# Spin asymmetries in lepton-proton and proton-proton diffractive reactions

S.V.Goloskokov,

BLTP, JINR, Dubna, Russia [1]

**Abstract**

It is shown that the longitudinal double spin asymmetry $A_{ll}$ in polarized diffractive $Q\bar{Q}$ production depends strongly on the spin structure of the quark-pomeron vertex. Relevant experiments will be possible at HERA with a polarized proton beam.

Diffractive production of high $p_t$ jets has been observed experimentally in hadron-hadron collisions [1] and in deep inelastic lepton-proton scattering [2]. Such processes where a proton stays intact or becomes a low mass state are determined at high energies by the pomeron exchange. These reactions with high $p_t$ jets can be interpreted as the observation of the partonic structure of a pomeron [3]. In what follows we shall suppose that the observed effects can be predominated by the quark structure of the pomeron [4, 5].

The pomeron is a vacuum $t$-channel exchange that contributes to high-energy diffractive reactions. There are two well-known models that are used to describe the pomeron contribution. First is the nonperturbative two-gluon exchange model [6, 7], and second is the BFKL pomeron [8] based on the summation of leading perturbative logarithms. Such a "bare" pomeron exchange leads to the mainly imaginary scattering amplitude with a simple quark-pomeron vertex structure

$$V_{qq\!P}^\mu \sim \gamma^\mu. \tag{1}$$

Then spin-flip effects are extremely small and in some sense in experiments [1, 2] the spin-average distributions inside the pomeron have been studied.

However, the spin structure of the quark-pomeron vertex may be not so simple. Separation of low and high-energy contributions in the pomeron exchange leads to factorization of the quark-proton scattering amplitude into the high-energy spinless "bare" pomeron and low-energy spin-dependent parts - quark-pomeron and hadron-pomeron vertices [9]. As a result, the pomeron contribution to the quark-proton amplitude looks as follows

$$T(s,t) = I\!\!P(s,t) V_{qq\!P}^\mu \otimes V_\mu^{pp\!P}, \tag{2}$$

where $I\!\!P$ is a "bare" pomeron contribution, $V_{qq\!P}$ and $V^{pp\!P}$ are the quark-pomeron and proton-pomeron vertices, respectively. The quark-pomeron vertex has been calculated perturbatively [9]. It has a form

$$V_{qq\!P}^\mu(k,r) = \gamma_\mu u_0(r) + 2m k_\mu u_1(r) + 2 k_\mu \slashed{k} u_2(r) + i u_3(r) \epsilon^{\mu\alpha\beta\rho} k_\alpha r_\beta \gamma_\rho \gamma_5 + i m u_4(r) \sigma^{\mu\alpha} r_\alpha. \tag{3}$$

---

[1] Email: goloskkv@thsun1.jinr.dubna.su

Here $k$ is a quark momentum, $r$ is a momentum transfer. The quantities $u_i(r)$ in (3) are the vertex functions. Note that the structure of the quark-pomeron vertex function (3) is drastically different from the "bare" pomeron. Really, only the term proportional to $\gamma_\mu$ corresponds to the standard pomeron vertex (1) which reflects the well-known approximation that the spinless quark-pomeron coupling is like a $C = +1$ isoscalar photon. The terms $u_1(r) - u_4(r)$ lead to the spin-flip in the quark-pomeron vertex in contrast to the term proportional to $u_0(r)$. The functions $u_1(r) \div u_4(r)$ at large $r^2$ are not very small [10]. Note that the phenomenological $V^\mu_{qq\mathbb{P}}$ vertex with $u_0$ and $u_1$ terms was proposed in [11].

The proton-pomeron vertex has a form [12]

$$V^\mu_{pp\mathbb{P}}(p,r) = mp_\mu A(r) + \gamma_\mu B(r) \qquad (4)$$

The quantity $A$ in (4) determines the transverse polarization and $B$ contributes to the longitudinal asymmetry. The ratio $m^2 A/B$ is about 0.2 [12].

Thus, the pomeron vertices have a complicated spin structure. This should modify different spin asymmetries and lead to new effects in high energy diffractive reactions that can be measured in future spin experiments at HERA, for example.

One of the simplest way to test the quark-pomeron vertex is to study the $Q\bar{Q}$ production in diffractive reactions. This sort of reaction has been investigated by different authors for unpolarized particles (see [4, 13, 14], e.g.).

In this report we shall discuss the longitudinal double spin asymmetries in polarized $p\uparrow p\uparrow \to p + Q\bar{Q} + X$ and $l\uparrow p\uparrow \to l + p + Q\bar{Q}$ diffractive reactions. Such experiments will be possible in future if the proton beam at HERA is polarized. Single transverse spin asymmetry will be discussed in an other talk.

Let us analyse the $Q\bar{Q}$ production in diffractive $pp$ scattering (see fig.1). It was shown in our previous estimations [15] that $A_{ll}$ asymmetry in this case can reach $10 \div 12\%$. Standard kinematical variables look as follows

$$s = (p_i + p)^2, \ t = r^2 = (p - p')^2, \ x_p = \frac{p_i(p - p')}{p_i p}. \qquad (5)$$

Here $p_i$ and $p$ are initial proton momenta, $p'$ is a momentum of recoiled hadron, $r$ is a momentum transfer at the pomeron vertex and $x_p$ is a part of momenta $p$ carried off by the pomeron. The diagram of Fig.1 is important at small $x_p$ that leads to a small invariant mass in the $Q\bar{Q}$ system $M_x^2 \sim \bar{y} x_p s$. The diagram with triple pomeron vertex must be considered for large $M_x^2$. Then we shall see more than two high $p_t$ jets.

The pomeron will be a leading contribution when the energy $s_p$ in the $Qp$ system is sufficiently large. One can obtain

$$s_p = (k + p)^2 \geq \frac{|t| + M_Q^2}{x_p}. \qquad (6)$$

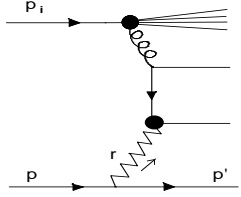

Figure 1: The diffractive $Q\bar{Q}$ production in $pp$ reaction.

So, $s_p$ is sufficiently large when $|t|$ is a few $GeV^2$ and $x_p \sim 0.1 \div 0.2$. For such $|t|$ the perturbative QCD can be used to calculate the spin structure of quark-pomeron vertex.

We shall calculate the longitudinal double spin asymmetry determined by the relation

$$A_{ll} = \frac{\Delta\sigma}{\sigma} = \frac{\sigma(\overset{\rightarrow}{\Leftarrow}) - \sigma(\overset{\rightarrow}{\Rightarrow})}{\sigma(\overset{\rightarrow}{\Rightarrow}) + \sigma(\overset{\rightarrow}{\Leftarrow})}. \tag{7}$$

In the longitudinal asymmetry case only the $\gamma_\mu B(r)$ term at the proton-pomeron vertex (4) contributes.

In calculations we shall use the hypothesis that the pomeron couples to a single quark (or an antiquark) for simplicity. The integration over the all $Q\bar{Q}$ phase space will be performed. For a planar loop we find

$$\sigma(\Delta\sigma) = F(I\!P)\frac{c}{x_p}\int_{y_0}^{1} dy g(y)(\Delta g(y)) \int_0^{syx_p/4} \frac{d^2k_\perp N^{\sigma(\Delta\sigma)}(x_p, k_\perp^2, u_i, |t|)}{\sqrt{1 - 4k_\perp^2/syx_p}(k_\perp^2 + M_Q^2)^2}. \tag{8}$$

Here $g(\Delta g)$ are the gluon spin-average and spin structure function of the proton, $k_\perp$ is a transverse part of the momentum in the loop, $M_Q$ is a quark mass, $N^{\sigma(\Delta\sigma)}$ is a trace over the quark loop. In (8) $F(I\!P)$ is a function which contains "bare" pomeron contribution $I\!P(s,t)$ and pomeron-proton vertices $V^{ppI\!P}$ (see (2)). This, function is the same for $\sigma$ and $\Delta\sigma$ in (7) and cancels at $A_{ll}$. Similar forms were found for the nonplanar loop.

We find that $\sigma \propto 1/x_p^2$ at small $x_p$. This behaviour is associated usually with the pomeron flux factor for $\alpha_{I\!P}(0) = 1$. However, $\Delta\sigma$ is proportional to $\epsilon^{\mu\nu\alpha\beta}r_\beta... \propto x_p p$. Thus, additional $x_p$ appears and we find that $\Delta\sigma \propto 1/x_p$ at small $x_p$.

In calculations of the integrals (8) the off-mass-shell behaviour of the pomeron structure functions $u_i$ has been considered. It was found that the following forms can be used as a good approximation for all functions at not small $|t|$

$$u_i(k_\perp^2, |t|) = \frac{|t|}{|t| + k_\perp^2} u_i(0, |t|). \tag{9}$$

As a result, the convergence of integrals (8) over $k_\perp$ is improved.

We use the simple form of the $u_0(r)$ vertex function

$$u_0(r) = \frac{\mu_0^2}{\mu_0^2 + |t|}, \quad r^2 = |t|,$$

with $\mu_0 \sim 1 Gev$ introduced in [16]. The functions $u_1(r) \div u_4(r)$ at $|t| > 1 GeV^2$ were calculated in perturbative QCD [10].

The main contribution to $\Delta \sigma$ is proportional to the first moment of $\Delta g$

$$\Delta g = \int_0^1 dy \Delta g(y), \tag{10}$$

which is unknown now. However, the magnitude of $\Delta g$ can be large, $\Delta g \sim 3$. This large magnitude of $\Delta g$ is important in the explanation of the proton spin [17].

In calculation of $\sigma$ we use the simple form of the gluon structure function

$$g(y) = \frac{R}{y}(1-y)^5, \quad R = 3.$$

This form corresponds to the pomeron with $\alpha_{I\!P}(0) = 1$. Just the same approximation for the pomeron exchange has been used in calculations. The analysis can be made for the pomeron with $\alpha_{I\!P}(0) = 1 + \delta$ and more complicated structure functions but it does not change the results drastically.

The resulting asymmetry depends on the ratio

$$C_g = \frac{\Delta g}{R}. \tag{11}$$

If (10) is fulfilled, $C_g \sim 1$. This magnitude will be used in what follows.

For a standard form of the pomeron vertex (1) we find

$$A_{ll} = \frac{-2x_p(\ln \frac{|t|}{M_Q^2} - 3)}{\ln \frac{|t|}{M_Q^2}(2\ln \frac{sx_p}{4|t|} + \ln \frac{|t|}{M_Q^2})}. \tag{12}$$

For the pomeron vertex (3) the axial-like term $V^\mu(k,r) \propto u_3(r)\epsilon^{\mu\alpha\beta\rho}k_\alpha q_\beta \gamma_\rho \gamma_5$ is extremely important in asymmetry. The formula for asymmetry is more complicated in this case.

Our predictions for $A_{ll}$ asymmetry at $\sqrt{s} = 40 GeV$ and $x_p = 0.2$ for the standard quark-pomeron vertex ($\gamma_\mu$ term is taken into account in (3)) and the spin-dependent quark-pomeron vertex (3) are shown in Fig.2 for light quarks and in Fig.3 for the heavy (C) quark. It is easy to see that the obtained asymmetry strongly depends on the structure of the quark-pomeron vertex. For the spin-dependent quark-pomeron vertex, $A_{ll}$ asymmetry is smaller by factor 2 because $\sigma$ in (7) is larger in this case. This is connected with the contribution of other $u_i$ structures.

As it was mentioned above, the asymmetry in $pp$ polarized diffractive reactions depends on an unknown $\Delta g$ spin gluon structure function of the proton. To obtain more explicit results, let us study $Q\bar{Q}$ diffractive production in lepton-proton reaction. At small $x_p$ it is determined by a diagram similar to that shown in Fig.1 with the lepton and photon instead of the gluon structure function of proton in the upper part of Fig.1.

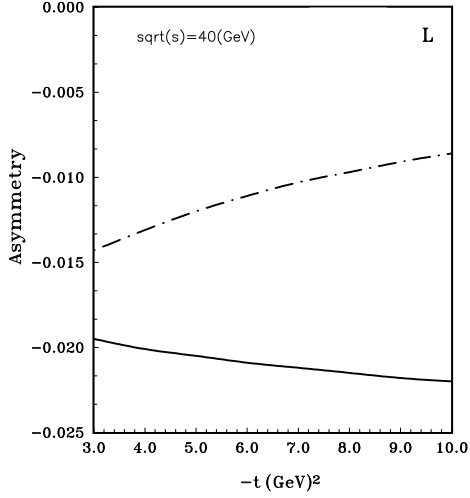 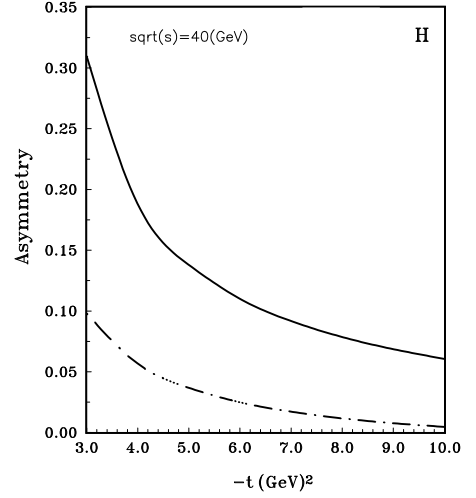

Fig.2  Fig.3

Figure 2: The $A_{ll}$ asymmetry of light quarks production. Solid line -for standard; dot-dashed line -for spin-dependent quark-pomeron vertex.

Figure 3: The $A_{ll}$ asymmetry of heavy (C) quarks production. Solid line -for standard; dot-dashed line -for spin-dependent quark-pomeron vertex.

The standard set of kinematical variables looks as follows [2]

$$s = (p_l + p)^2, \quad Q^2 = -q^2, \quad t = (p - p')^2$$
$$y = \frac{pq}{p_l p}, \quad x = \frac{Q^2}{2pq}, \quad \beta = \frac{Q^2}{2q(p - p')}, \quad x_p = \frac{q(p - p')}{qp}, \tag{13}$$

where $p_l, p'_l$ and $p, p'$ are initial and final lepton and proton momenta, respectively, $q = p_l - p'_l$.

The asymmetry is determined by formula (7). For a planar loop we find

$$\sigma(\Delta\sigma) = F(I\!P)\frac{c}{x_p}\int_0^{syx_p/4} \frac{d^2 k_\perp N^{\sigma(\Delta\sigma)}(x_p, k_\perp^2, u_i, |t|)}{\sqrt{1 - 4k_\perp^2/syx_p}(k_\perp^2 + M_Q^2)^2} \tag{14}$$

The main contributions to $A_{ll}$ asymmetry in the discussed region are determined by the $u_0$ and $u_3$ structures in (3). For a standard form of the pomeron vertex (1) we have

$$A_{ll} = -x_P y \frac{Q^2(2-y)(\ln(\frac{|t|}{m_Q^2}) - 3)}{[2|t|(1-y)\ln(\frac{Q^2}{|t|\beta}) + Q^2 y(2-y)\ln(\frac{|t|}{m_Q^2})]}. \tag{15}$$

The formulae for $\sigma$ and $\Delta\sigma$ for the pomeron vertex (3) can be found in [18]. Note that $\Delta\sigma$ is proportional to $Q^2$. As a result the asymmetry must increase with $Q^2$ (15).

Our predictions for $A_{ll}$ asymmetry for energy $\sqrt{s} = 300 GeV$ estimated from perturbative vertex functions for $y = 0.5$ and $x_p = 0.2$ for the standard quark pomeron vertex and the spin-dependent quark pomeron are shown in Fig. 4,5. In fig.4 the $|t|$ dependence of $A_{ll}$ for fixed $Q^2 = 10 GeV^2$ ($\beta = 0.25$) is shown. In fig.5 one can see the $Q^2$ dependence of $A_{ll}$ for fixed $|t| = 3 GeV^2$. The obtained asymmetry is not small and strongly depends on the spin structure of the quark-pomeron vertex. Asymmetry decreases with $|t|$ growing and increase with $Q^2$ growing.

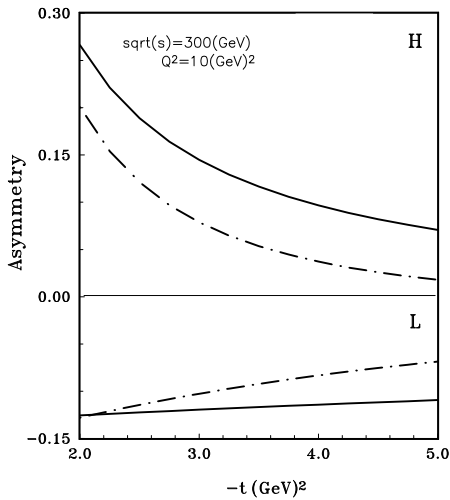 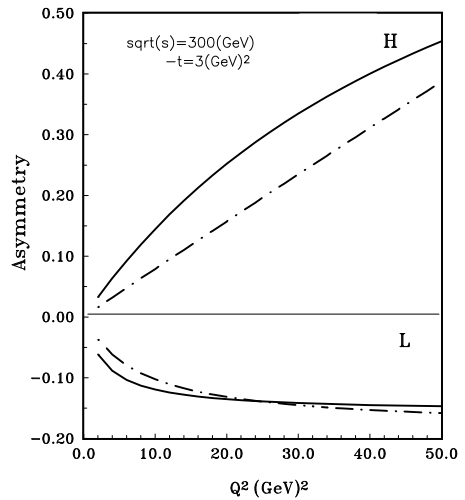

Fig.4  Fig.5

Figure 4: The $|t|$ dependence of $A_{ll}$ asymmetry of light and heavy (C) quarks production at fixed $Q^2 = 10 GeV^2$. Solid line -for standard; dot-dashed line -for spin-dependent quark-pomeron vertex.

Figure 5: The $Q^2$ dependence of $A_{ll}$ asymmetry of light and heavy (C) quarks production at fixed $|t| = 3 GeV^2$. Solid line -for standard; dot-dashed line -for spin-dependent quark-pomeron vertex.

The estimations show that total integrated cross section of light quark production in $lp$ reaction is about $0.2 \div 0.1$ nb [4, 14]. Our calculation shows that the cross section for $C$ quark production has to be smaller by factor $3 \div 10$.

The obtained asymmetries in lepton-proton and proton-proton reactions have some common properties

- The asymmetry for heavy quark production is sufficiently large and positive.

- Asymmetry is opposite in sign for light and heavy quarks. This is determined by a factor like $(ln(|t|/M_Q^2) - 3)$ that appears in the asymmetry. It is different in sign for $M_Q \sim 0.005 GeV$ and $M_Q \sim 1.5 GeV$ for the investigated momenta transfer.

- Asymmetry decreases with energy only logarithmically $A_{ll} \sim 1/ln(sx_p/(4|t|))$.

- Asymmetry is equal to zero at $x_p = 0$. So, it is better to study it at $x_p = 0.1 \div 0.2$

- Obtained asymmetry strongly depends on the structure of the quark-pomeron vertex.

- The axial-like term $V^\mu(k,r) \propto u_3(r)\epsilon^{\mu\alpha\beta\rho}k_\alpha q_\beta \gamma_\rho \gamma_5$ is extremely important in asymmetry. This term is proportional to momenta transfer $r$ ($r^2 = |t|$) and can be measured only for the pomeron contribution with nonzero momenta transfer. Then the detection of a final proton is extremely important, otherwise, the integration over $|t|$ can be performed (see [2] e.g.).

The discussed spin-dependent contributions to the quark-pomeron and hadron-pomeron vertex functions modify different spin asymmetries and lead to new effects in high energy diffractive reactions which can be measured in spin experiments at future accelerators.

To summarize, we have presented in this letter the perturbative QCD analysis of the longitudinal double spin asymmetry in diffractive 2-jet production in $lp$ and $pp$ processes. The model prediction shows that the $A_{LL}$ asymmetry can be studied and the information about the spin structure of the quark-pomeron vertex can be extracted. It should be emphasized that the obtained spin effects are completely determined at fixed momenta transfer by the large-distance contributions in quark (gluon) loops. So, they have a nonperturbative character. The investigation of spin effects in diffractive reactions is an important test of the spin sector of QCD at large distances.

This work was supported in part by the Russian Fond of Fundamental Research, Grant 94-02-04616 and Heisenberg-Landau Grant.

# References

[1] A.Brandt, Phys.Lett., 1992, **B297**, 417.

[2] T.Ahmed et al, Preprint DESY 95-36, DESY 1995.

[3] G.Ingelman, P.E.Schlein, Phys.Lett., 1985, **B152**, 256.

[4] A.Donnachie, P.V.Landshoff, Phys.Lett., 1992, **B285**, 172.

[5] J.C.Collins, L.Frankfurt, M.Strikman, Phys.Lett., 1993, **B307**, 161.

[6] F.E.Low, Phys.Rev., 1975, **D12**, 163;
    S.Nussinov, Phys.Rev.Lett., 1975, **34**, 1286.

[7] P.V.Landshoff, O. Nachtmann, Z.Phys., 1987, **C35**, 405.

[8] E.A.Kuraev, L.N.Lipatov, V.S.Fadin, Sov.Phys. JETP, 1976, **44**, 443;
Y.Y.Balitsky,L.N.Lipatov, Sov.J.Nucl.Phys., 1978, **28**, 822.

[9] S.V.Goloskokov, Phys.Lett., 1993, **B315**, 459.

[10] S.V.Goloskokov, O.V. Selyugin, Yad. Fiz., 1994, **57**, 727.

[11] J.Klenner, A.Schafer, W.Greiner, E-print hep-ph 9409451.

[12] S.V.Goloskokov, S.P.Kuleshov, O.V.Selyugin, Particles and Nucl., 1987, **18**, 39.

[13] E.L.Berger et al., Nucl.Phys., 1987, **B286**, 704;
V.Del. Duca, E-print hep-ph 9503335.

[14] M.G.Ryskin, S.Yu.Sivoklokov, A.Solano. In Proc. of Int. Conf. on Elastic and Diffractive Scattering, Ed. by H.M.Fried, K.Kang, C-I Tang, World Sci, 1993, p.181.

[15] S.V.Goloskokov, O.V.Selyugin, JINR Rapid Comm. 2-94, HEP-PH/9403337.

[16] A.Donnachie, P.V.Landshoff, Nucl.Phys., 1989, **B311** 509.

[17] A.V.Efremov, J.Soffer, O.V.Teryaev, Nucl.Phys., 1990, **B396**,97;
M.Anselmino, A.V.Efremov, E.Leader, CERN-TH 7216/1994, CERN 1994.

[18] S.V.Goloskokov, E-print hep-ph 9506347.